\documentclass[titlepage]{csetr}

\usepackage{graphicx}
\usepackage{subfigure}

\renewcommand{\and}{\hspace{.5cm}}

\trnumber{1019}
\trdate{November 2010}

\title{%
  Extending SPARQL to Support Entity Grouping\\ and Path Queries
}
\author{%
  Seyed-Mehdi-Reza Beheshti$^1$ \and %
  Sherif Sakr$^1$ \\ %
  Boualem Benatallah$^1$ \and %
  Hamid Reza Motahari-Nezhad$^2$\\[2em]
  $^1\, $University of New South Wales\\ Sydney 2052, Australia \\%
  \email{\{sbeheshti,ssakr,boualem\}@cse.unsw.edu.au}\\ \\
  $^2\,$HP Labs Palo Alto\\ CA 94304, USA \\%
  \email{hamid.motahari@hp.com}\\[3cm]
}

\date{}
\begin{document}
\maketitle

\begin{abstract}
The ability to efficiently find relevant subgraphs and paths in a large graph to a given query is important in many applications including scientific data analysis, social networks, and business intelligence. Currently, there is little support and no efficient approaches for expressing and executing such queries. This paper proposes a data model and a query language to address this problem. The contributions include supporting the construction and selection of: (i) folder nodes, representing a set of related entities, and (ii) path nodes, representing a set of paths in which a path is the transitive relationship of two or more entities in the graph. Folders and paths can be stored and used for future queries. We introduce FPSPARQL which is an extension of the SPARQL supporting folder and path nodes. We have implemented a query engine that supports FPSPARQL and the evaluation results shows its viability and efficiency for querying large graph datasets.\end{abstract}

\section{Introduction}

Graph is a generic structure for representing data in many domains including business intelligence, scientific data analysis, provenance, bibliographic networks, and social networking~\cite{GraphBook}. With enormous amount of data available, the resultant graphs are very large. An example of this is the case of Web and social network graphs, which may contain millions of nodes. The need for efficient approaches for querying and analyzing these graphs is emergent. In particular, manipulating, querying, and analyzing graphs to discover new knowledge is of high interest~\cite{GraphBook}.

Among various types of queries on data graphs, those which return a graph, a set of subgraphs or a set of paths in the large graph are gaining attention. One such query is finding the influence graph of a paper in a bibliographic graph through analyzing the citation of the paper in the graph. As another example, we may want to find a set of related activities in a business process graph, as an informal description of the process may be available in the form of a process graph. There is a need for graph representation models and efficient approaches for expressing and executing these types of queries.

Among languages for querying graphs, SPARQL is a declarative query language, an official W3C standard and widely used for querying and extracting information from directed-labeled RDF graphs~\cite{SPARQL}. It is based on a powerful graph matching mechanism that allows binding variables to components in the input graph and supports conjunctions and disjunctions of triple patterns. In addition, operators akin to relational joins, unions, selections, and projections can be combined to build more expressive queries. However, SPARQL does not support the construction and retrieval of subgraphs. Also paths are not first class objects in SPARQL~\cite{SPARQL,ProvQuery}.

Addressing this problem is challenging, as there is a need for a comprehensive, scalable, and efficient query language for graph analysis. The language should be native to graphs, general enough to meet the heterogeneous nature of real world data, and declarative~\cite{GraphBook}. In this paper, we present an approach for representing and querying graphs. The main contributions of the paper are as follows:

\begin{itemize}
\item We propose a graph data model that supports structured and unstructured entities, and introduces \textit{folder} and \emph{path} nodes as first class abstractions. A folder node contains a collection of related entities, and a path node represent the results of a query that consists of one or more paths in the graph (a path is defined based on a transitive relationship between two entities).

\item We define the FPSPARQL query language, a Folder-Path enabled extension of the SPARQL, to manipulate and query entities, and folder and path nodes.

\item We describe the implementation of a query engine supporting FPSPARQL, and evaluate our query engine over large datasets.
\end{itemize}

The remainder of this paper is organized as follows: We present the data model in section \ref{DataModel}, and illustrate the manipulation part of the data model in section \ref{DataManipulation}. In section \ref{QueryEngine} we discuss the query engine implementation. Section \ref{ExperimentalEvaluation} presents a motivating scenario as an experiment and evaluates the proposed query engine. Section \ref{RelatedWork} presents related work. Finally, we conclude the paper with a prospect on future work in Section \ref{Conclusion}.

\section{Graph Data Model, Folders, and Paths}
\label{DataModel}

\begin{figure}[t]
\centering
    \includegraphics[width=1\textwidth]{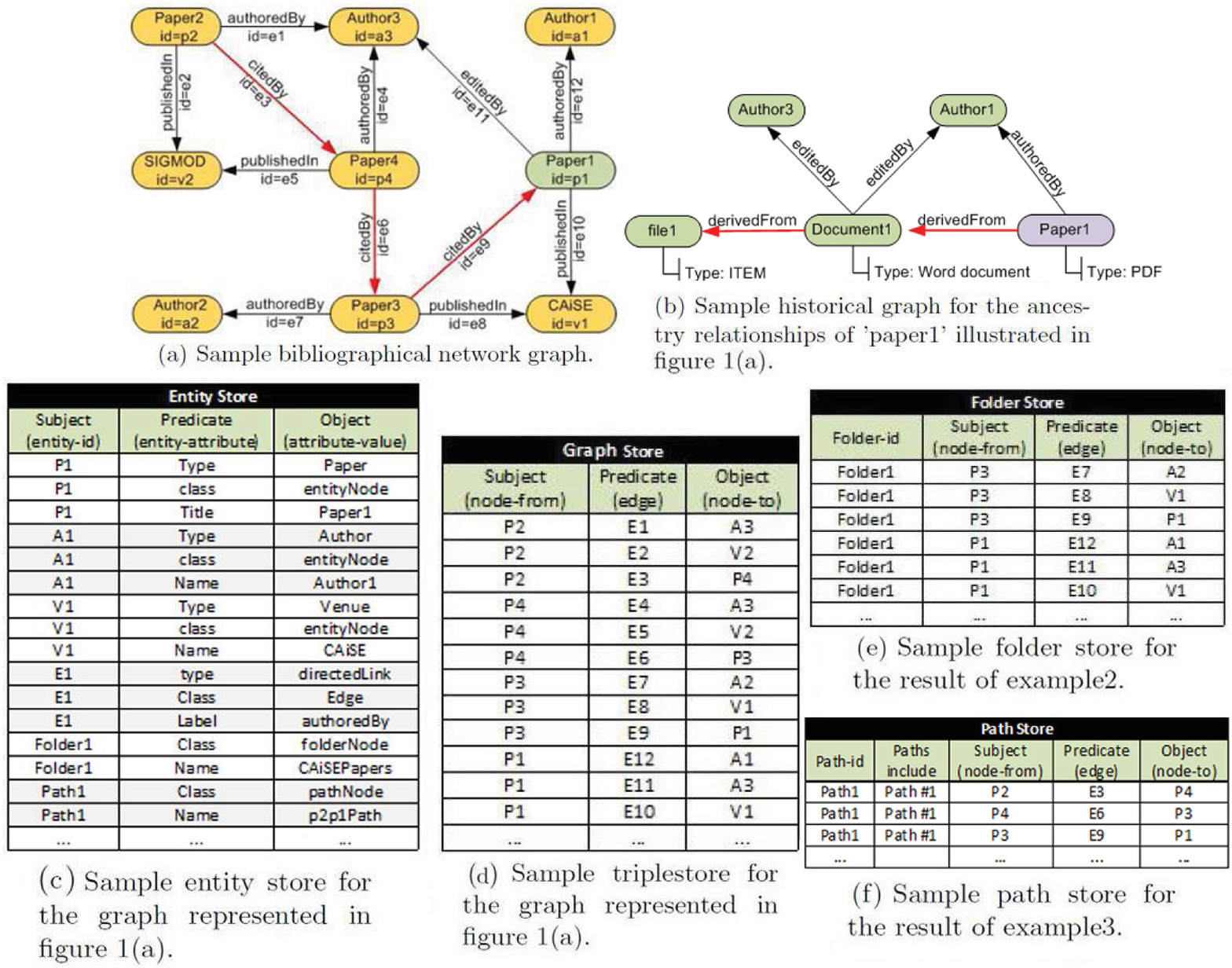}
    \caption{Representation of graph, Folder, and Path.} \label{fig:exampleEFPN}
\end{figure}

We define a graph data model for organizing a set of entities as graph nodes and entity relationships as edges of the graph. An entity is a data object that exists separately and has a unique identity. This data model supports: (i) structured and unstructured entities; (ii) folder nodes, which contain entity collections. A folder node represent the results of a query that returns a collection of related entities; and (iii) path nodes, which refer to one or more paths in the graph, which are the result of a query. A path is the result of the transitive relationship between two entities. Entities and relationships are represented as a directed graph $G =(V, E)$ where $V$ is a set of nodes representing entities, folder or path nodes, and $E$ is a set of directed edges representing relationships between nodes. 

\subsection{Entities}
Entities could be structured or unstructured. Structured entities are instances of entity types. An entity type consists of a set of attributes. Unstructured entities, are also described by a set of attributes but may not conform to an entity type. This entity model offers flexibility when types are unknown and take advantage of structure when types are known. For the sake of simplicity, we assume all unstructured entities are instances of a generic type called \emph{ITEM}. ITEM is similar to \emph{generic table} in \cite{OneTableStoresAll}. We store entities in the \emph{entity store}.\\

\noindent \emph{Example 1.} Consider the bibliographical graph in Figure 1(a). In this graph we have entity types such as author, paper and venue. The graph in Figure 1(b) illustrates the creation (i.e. ancestry relationships) of 'paper1'. 'paper1' and 'document1' are structured entities. 'file1' is an unstructured entity with unknown entity type. 
The sample entity store in Figure 1(c) contains all the entities in this graph. The graph store in Figure 1(d), contains the directed links between entities.

\subsection{Relationships}
A relationship is a directed link between a pair of entities, which is associated with a predicate defined on the  attributes of entities that characterizes the relationship. A relationship can be \emph{explicit}, such as \emph{authoredBy} in 'paper \emph{authoredBy} author' in a bibliographical network. Also a relationship can be \emph{implicit}, such as a relationship between an entity and a larger (composite) entity that can be inferred from the nodes.

\subsection{Folder nodes}
A folder node contains a set of entities that are related to each other. In other words, the set of entities in a folder node is the result of a given query that require grouping graph entities in a certain way. A folder node creates a higher level node that other queries could be executed on top of it. Folders can be nested, i.e., a folder can be a member of another folder node, to allow creating and querying folders with relationships at higher levels of abstraction. A folder may have a set of attributes that describes it. A folder node is added to the graph and can be stored in the \emph{folder store} to enable reuse of the query results for frequent or recurrent queries.\\

\noindent \emph{Example 2.} As an example of a relationship, let us consider a correlation condition for two entities defined as a binary predicate over attributes of the entities. We call two entities correlated if the predicate is evaluated to true. Consider the correlation condition \emph{x.venue='CAiSE'} where $x$ is an instance of type \emph{paper}. This query, groups set of papers published in 'CAiSE' conference. As illustrates in the Figure 1(a) the result of this query is the set \{'paper1','paper3'\}. We add a folder node to the original graph, and store the result of this query in the folder store (Figure 1(e)). For this purpose, we filter all the tuples in the graph store (Figure 1(d)) whose column 'node-from' is 'paper1' or 'paper3'. 
Properties of this folder will be stored in the entity store (Figure 1(c)). In the folder store, the nodes under the column 'subject' are the members of this folder.

\subsection{Path nodes}
A path is a transitive relationship between two entities showing the sequence of edges from the start entity to the end. This relationship can be codified using regular expressions~\cite{RE_path,CodebookPath} in which alphabets are the nodes and edges from the graph. We define a path node for each query that results in a set of paths. We use existing reachability approaches to verify whether an entity is reachable from another entity in the graph. Some reachability approaches (e.g. all-pairs shortest path~\cite{CodebookPath}) report all possible paths between two entities. We define a path node as a triple of $(V_{start},V_{end},RE)$ in which $V_{start}$ is the start node, $V_{end}$ is the end node and a regular expression $RE$. We store all paths of a path node in the \emph{path store}.

For example, in a bibliographic graph, one possible query that results in a set of paths in the graph is ``find all conferences for papers citing a given paper''. Such a query will help to understand which conferences cite papers from a given conference. The details of such a query is a set of paths from the current paper to the publication venue of papers citing the given paper. In cases, where the second entity of a target path query is not given, the query requires a maximum length to limit the search for matching end entities within that maximum length from the start entity.\\

\noindent \emph{Example 3.} Consider the bibliographical network presented in Figure 1(a). Assume we are interested in finding occurrences of following pattern: 'paper2' cited by 'paper1' possibly indirectly (follow the red edges in the figure). This path can be written as regular expressions, starting with "paper (citedBy paper)+". The plus sign indicates that there is one or more of the preceding element. The result of this example stored in a sample path store presented in Figure 1(f).\\

\noindent \emph{Example 4.} Consider the historical graph presented in Figure 1(b). The ancestry relationships found in provenance\footnote{Provenance \cite{Moreau:OPM} is the process of recording events happening in digital environments which generates the documented history of information items' creation.} form a directed graph, i.e. historical graph. When an object A is found to have been derived from some other object B, we say that there is an ancestry path between A and B \cite{ProvQuery}. Figure 1(b) illustrates the ancestry path between 'paper1' and 'file1' (follow the red edges in the figure). Ancestry paths through historical graphs form the basis of many provenance queries.

\section{Data Manipulation}
\label{DataManipulation}
Our graph-based data model and query requirements are very similar to those in the SPARQL query language \cite{SPARQL}. Thus, we decided to base our language on SPARQL. We support two levels of queries in FPSPARQL: (i) Graph-level Queries: at this level we use SPARQL to query  graphs; and (ii) Node-level Queries: at this level we propose an extension of SPARQL to construct and query folder nodes and path nodes.

\subsection{Graph-level Queries}

SPARQL is an RDF query language, standardized by the World Wide Web Consortium, for semantic web. SPARQL contains capabilities for querying required and optional graph patterns along with their conjunctions and disjunctions. SPARQL also supports extensible value testing and constraining queries. The results of SPARQL queries can be results sets or RDF graphs. A basic SPARQL query has the form:

\begin{verbatim}
 select ?variable1 ?variable2 ...
 where {
 pattern1. pattern2. ...
 }
\end{verbatim}

Each pattern consists of \emph{subject}, \emph{predicate} and \emph{object}, and each of these can be either a variable or a literal. The query specifies the known literals and leaves the unknowns as variables. 
To answer a query we need to find all possible variable bindings that satisfy the given patterns. We use the '@' symbol for representing attribute edges and distinguishing them from the relationship edges between graph nodes. Example 5 presents a sample graph-level query.\\

\begin{figure}[t]
\centering
\subfigure[Sample RDF Graph.] {
    \label{fig:sparqlRDF}
    \includegraphics[width=0.55\textwidth]{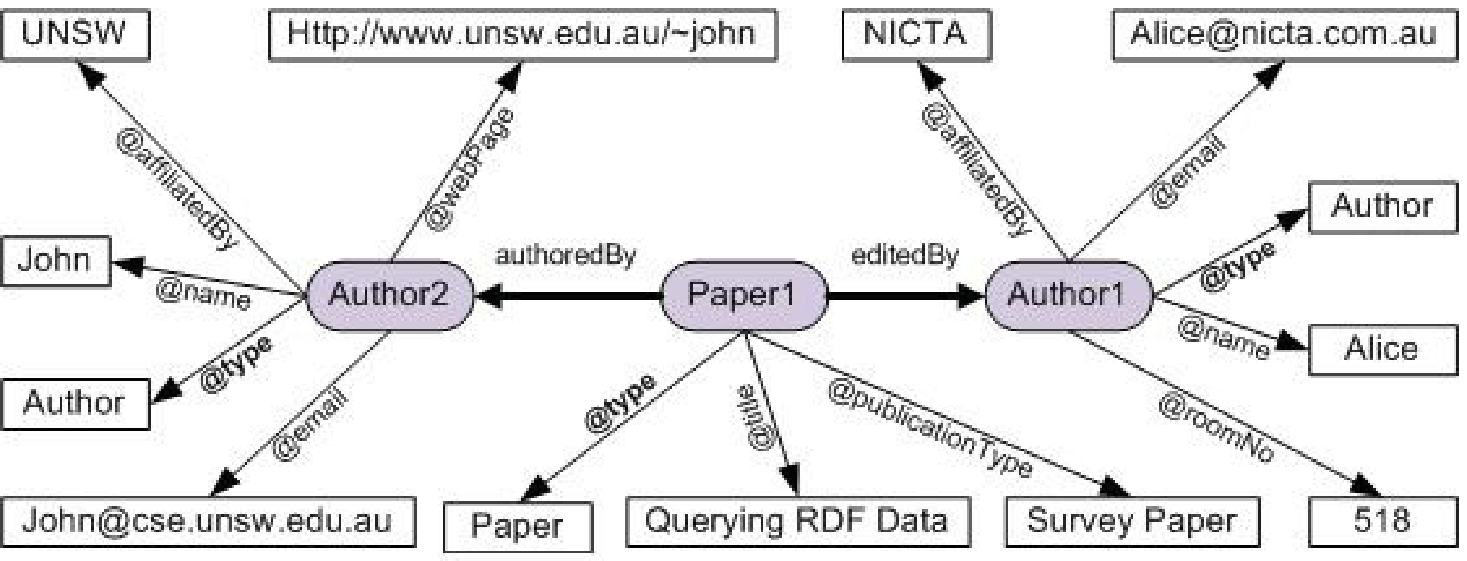}
}
\hspace{0.2cm}
\subfigure[Sample SPARQL query.] {
    \label{fig:sparqlQuery}
    \includegraphics[width=0.35\textwidth]{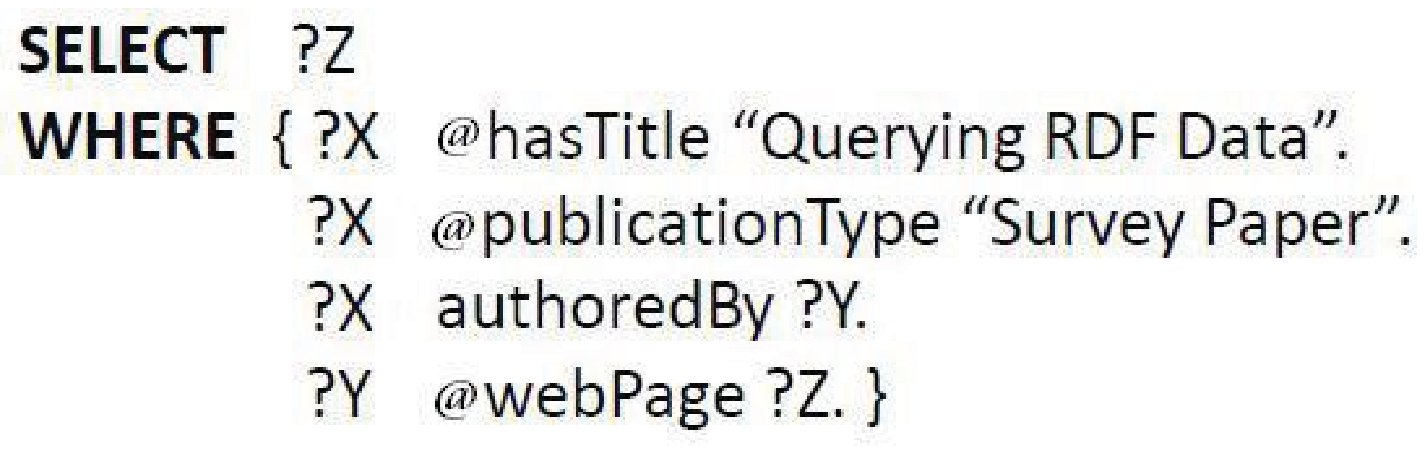}
}
\caption{Representation of the graph-level query proposed in example 4.} \label{fig:example4}
\end{figure}

\noindent \emph{Example 5.} Figure~\ref{fig:sparqlQuery} depicts a sample SPARQL query over the sample RDF graph of Figure~\ref{fig:sparqlRDF} to retrieve the web page information of the author of a book chapter with the title "Querying RDF Data".

\subsection{Node-level Queries}

Standard SPARQL querying mechanisms is not enough to support querying needs of FPSPARQL and its data model. In particular, SPARQL does not support folder nodes and querying them natively and such queries needs to be applied to the whole graph. In addition, querying the result of a previous query becomes complex and cumbersome, at best. Also path nodes are not first class objects in SPARQL \cite{CodebookPath,ProvQuery}. We extend SPARQL to support node-level queries to satisfy specific querying needs of our data model. Node-level queries in FPSPARQL include two special constructs: (a) CONSTRUCT queries: used for constructing folder and path nodes, and (b) APPLY queries: used to ease the requirement of applying queries on folder and path nodes.

\subsubsection{Folder Node Construction.} To construct a folder node, we introduce the $FCONSTRUCT$ command. This command is used to group a set of related entities or folders. A basic folder node construction query looks like this:\\\\

\begin{verbatim}
 fconstruct <Folder_Node Name>
 [ select ?var1 ?var2 ... |
 (Folder_Node1 Name, Folder_Node2 Name, ...) ]
 where { pattern1. pattern2. ... }
\end{verbatim}

A query can be used to define a new folder node by listing folder node name and entity definitions in the \emph{fconstruct} and
\emph{select} statements, respectively. Example 6 represents such a query. Also a folder node can be defined to group a set of folder nodes. A simple example of such a query represented in example~7. A set of user defined attributes for this folder can be defined in the \emph{where} statement.\\

\noindent \emph{Example 6.} Construct a folder node (and name it CAiSEPapers) for the query represented in example 2.

\begin{verbatim}
 fconstruct CAiSEPapers as ?fn
 select ?p
 where {
 ?fn @description 'set of ...'.
 ?p @type paper.
 ?p publishedIn 'CAiSE'.
 }
\end{verbatim}

In this example the variable ?p represent the papers published in 'CAiSE' conference. The variable ?fn represent the folder node to be constructed, i.e. 'CAiSEPapers'. This folder node has a user defined attribute called 'description'.\\

\noindent \emph{Example 7.} Consider 3 folder nodes 'SIGMOD08','SIGMOD09', and 'SIGMOD10' each representing accepted papers in SIGMOD conference 2008, 2009, and 2010. Construct a folder node ( i.e. 'SIGMOD') to group these folder nodes.

\begin{verbatim}
 fconstruct SIGMOD as ?fn
 (SIGMOD08,SIGMOD09,SIGMOD10)
 where {
 ?fn @description 'set of related folder nodes'.
 }
\end{verbatim}

In this example the variable ?fn represent the folder node to be constructed, i.e. 'SIGMOD'. This folder node contains 3 folder nodes and has a user defined attribute 'description'. These folder nodes are hierarchically organized by \emph{part-of} (i.e. an implicit relationship) relationships.

\subsubsection{Path Node Construction.} We introduce the $PCONSTRUCT$ command to construct a path node. This command is used to discover transitive relationships between two entities and store it under a path node name. In general a basic path node construction query looks like this:\\\\

\begin{verbatim}
 pconstruct <Path_Node Name>
 (Start Node,End Node,Regular Expression)
 where {
 pattern1. pattern2. ...
 }
\end{verbatim}

A regular expressions can be used to define a transitive relationship between two entities, i.e. starting node and ending node. Attributes of starting node, ending node, and regular expressions alphabets (i.e. graph nodes and edges) can be defined in the \emph{where} statement. Example 8 represents such a query.\\

\noindent \emph{Example 8.} Consider the bibliographical network presented in Figure 1(a). Construct a path node for the possible transitive relationship between 'paper2' and 'paper1', to analyze the citations of 'paper2'.

\begin{verbatim}
 pconstruct p2p1Path
 (?startNode,?endNode,(?e ?n)* ?citedByEdge (?n ?e)*)
 where {
 ?startNode @id p2.
 ?endNode @id p1.
 ?n @isA entityNode.
 ?e @isA edge.
 ?citedByEdge @isA edge.
 ?citedByEdge @label citedBy.
 }
\end{verbatim}

In this example ?startNode denotes 'paper2' and ?endNode denotes 'paper1'. Respectively, ?e and ?n denotes any edges and nodes in the transitive relationship between 'paper2' and 'paper1'. And ?citedByEdge denotes an edge labeled 'citedBy' in the path node. The \emph{isA} attribute denotes the \emph{class} attribute of the entities (see Figure 1(c)). In the regular expression, parentheses are used to define the scope and precedence of the operators and the asterisk indicates there are zero or more of the preceding element. This regular expressions matches the path node (i.e. p2p1Path) 'paper2 citedBy paper4 citedBy paper3 citedBy paper1' (follow the red edges in the Figure 1(a)).

\subsubsection{Folder Node Queries.} We introduce the $APPLY$ command to retrieve information, i.e. by applying queries, from the underlying folder nodes. These queries can apply on one folder node or the composition of several folder nodes. Our model supports the standard set operations (union, intersect, and minus) to apply queries on the composition of several folder nodes. In general, a basic folder node query looks like this:

\begin{verbatim}
 [Folder Node | (Composition of Folder Nodes)] APPLY (
 select ?variable1 ?variable2 ...
 where {
 pattern1. pattern2. ...
 })
\end{verbatim}

A graph-level query can be applied on a folder node (see example 9) or composition of folder nodes (see example 10) by listing folder node or composition of folder nodes before \emph{apply} command, and placing the query in parenthesis after \emph{apply} command.\\

\noindent \emph{Example 9.} Consider the folder node \emph{CAiSEPapers} constructed in example~6. We are interested in applying the query ``retrieve the papers which authored by \emph{author1}'' on this folder node.

\begin{verbatim}
 (CAiSEPapers) apply(
 select ?p
 where {
 ?p @type paper.
 ?p authoredBy ?a.
 ?a @type author.
 ?a @name 'author1'.
 })
\end{verbatim}

In this example ?p denotes papers that fall inside \emph{CAiSEPapers} folder node, and the query will apply on these papers. The result will be papers published in 'CAiSE' conference which authored by 'author1'.\\

\noindent \emph{Example 10.} Consider that we have constructed two folder nodes \emph{CAiSEPapers} (set of papers published in 'CAiSE' conference) and \emph{SIGMODPapers} (set of papers published in 'SIGMOD' conference). We are interested in retrieving the papers that published in 'SIGMOD' or 'CAiSE' which authored by 'author1'.

\begin{verbatim}
 (CAiSEPapers union SIGMODPapers) apply (
 select ?p
 where {
 ?p @type paper.
 ?p authoredBy ?a.
 ?a @type author.
 ?a @name 'author1'.
 })
\end{verbatim}

In this example ?p denotes papers that fall inside both \emph{CAiSEPapers} and \emph{SIGMODPapers} folder nodes. The query ``retrieve the papers that authored by \emph{author1}'' will apply on these papers.

\subsubsection{Path Node Queries.} This type of query is used to retrieve information, i.e. by applying queries, from the underlying path node. Path node queries are similar to folder node queries and use the same command, i.e. $APPLY$ command. In general, a basic path node query looks like this:

\begin{verbatim}
 Path_Node_Name APPLY (
 select ?variable1 ?variable2 ...
 where {
 pattern1. pattern2. ...
 })
\end{verbatim}

A graph-level query can be applied on a path node by listing path node name before apply command, and placing the query in parenthesis after apply command. Example 11 represents a simple example of such query.\\

\noindent \emph{Example 11.} Consider the path node \emph{p2p1Path} constructed in example 8. We are interested to find papers in the transitive relationship between 'paper2' and 'paper1', that have the keyword 'SQL' in their titles.

\begin{verbatim}
 (p2p1Path) apply (
 select ?p
 where {
 ?p @type paper.
 ?p @title ?t.
 Filter regex(?t,"SQL").
 })
\end{verbatim}

In this example ?p denotes papers that fall inside \emph{p2p1Path} path node. The query ``retrieve the papers that have the keyword 'SQL' in their titles'' will apply on these papers.

\section{Implementation}
\label{QueryEngine}

\begin{figure}
\centering
  \includegraphics[scale=0.7]{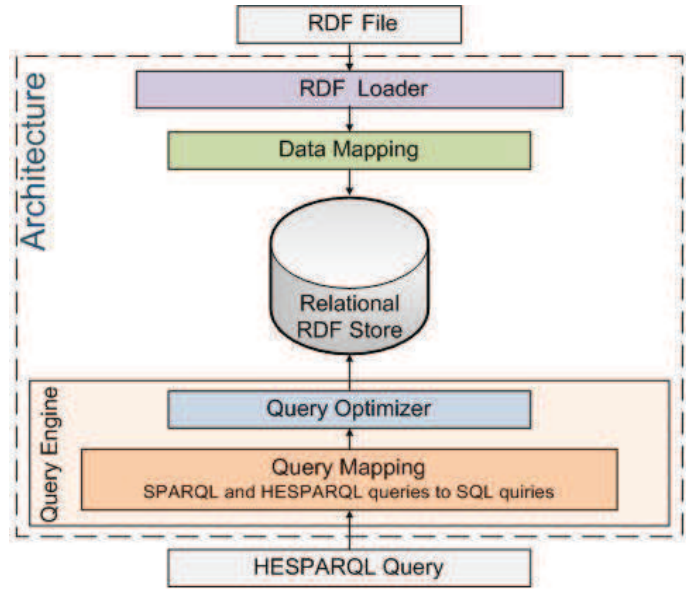}\\
  \caption{Engine Architecture.}\label{fig:QueryLanguageArchitecture}
\end{figure}

The simplest way to store a set of RDF statements is to use a relational database with a single table that includes columns for subject, property and object. While simple, this schema quickly hits scalability limitations~\cite{SherifSurvey1}. To avoid this we developed a relational RDF store including its three classification approaches~\cite{SherifSurvey1}: vertical (triple), property (n-ary), and horizontal (binary). The query engine is implemented in Java (J2EE) and consists of two main layers (Figure \ref{fig:QueryLanguageArchitecture}), data mapping and query engine.

\subsubsection{Data Mapping Layer.} This layer is responsible for creating data element mappings between semantic web technology (i.e. Resource Description Framework) and relational database schema. We developed a workload-independent physical design by developing a Loader algorithm. This algorithm is responsible for: (i) validating the input RDF; (ii) generating the relational representation of triple RDF store, for manipulating and querying entities, folders, and paths; and (iii) generating powerful indexing mechanisms.

\subsubsection{Query Engine.} 
\begin{figure}
\centering
  \includegraphics[scale=0.6]{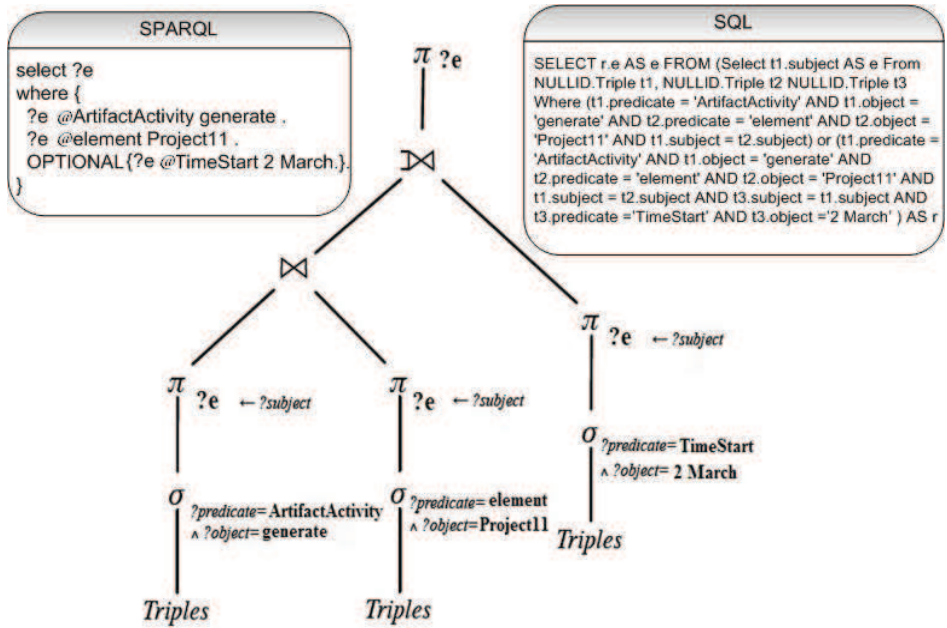}\\
  \caption{A SPARQL query, its translation into a relational operator tree, and its equivalent SQL query generated by our translation algorithm. }\label{fig:SPARQLTranslation}
\end{figure}

The query engine consists of two layers: query mapping and query optimizer. Query Mapping Layer consists of a FPSPARQL parser (for parsing FPSPARQL queries based upon the syntax of FPSPARQL) and a schema-independent FPS-PARQL-to-SQL translation algorithm. This algorithm consists of:
\begin{itemize}
  \item \emph{SPARQL-to-SQL translation algorithm.} We implemented a SPARQL-to-SQL translation algorithm based on the proposed relational algebra for SPARQL~\cite{SPARQLAlgebra} and semantics preserving SPARQL-to-SQL query translation~\cite{SPARQLSemantics}. This algorithm supports \emph{Aggregate} queries and \emph{Keyword Search} queries. Figure~\ref{fig:SPARQLTranslation} shows a SPARQL query, its translation into a relational operator tree, and its equivalent SQL query which is generated by this algorithm.
  \item \emph{Folder node construction and querying.} We use the relational representation of triple RDF store, to store, manipulate, and query folder nodes.
  \item \emph{Path node construction and querying.} To describe constraints on the path nodes, we reused the specification for regular expressions and filter expressions proposed in CSPARQL~\cite{RE_path}. We developed a regular expression processor which supports optional elements (?), loops (+,*), alternation (|), and grouping ((...))~\cite{CodebookPath}. We provide the ability to call external graph reachability algorithms (see section \ref{ApproximationQuality}) for path node queries.
\end{itemize}

To optimize the performance of queries, we developed four optimization techniques proposed in \cite{RDFProv,SherifSurvey1,SPARQLSemantics}: (i) selection of queries with specified varying degrees of structure and spanning keyword queries; (ii) selection of the smallest table to query based on the type information of an instance; (iii) elimination of redundancies in basic graph pattern based on the semantics of the patterns and database schema; and (iv) create separate tables (property tables) for subjects that tend to have common properties to reduce the self-join problem.

\section{Experimental Evaluation} \label{ExperimentalEvaluation}

In this section we provide an experimental evaluation of FPSPARQL query engine. We utilized IBM DB2 as a back-end database. All experiments were conducted on a HP system with a 2.67Ghz Core2 Quad processor, 4 GBytes of memory, and running a 64-bit Windows 7. We give an overview of the evaluation example and datasets used in section \ref{MotivatingExample}. We compared our system with HyperGraphDB~\cite{HyperGraphDB} (an open-source graph database) and present query running time measurements in section \ref{QueryExecutionTime}. We provide the ability to call external graph reachability algorithms for path node queries. We discuss the quality of finding paths by different approaches in subsection \ref{ApproximationQuality}.

\subsection{Evaluation Example}
\label{MotivatingExample}

Our example falls inside business intelligence domain and comes from our experience on managing an online project-based course ``e-Enterprise Projects''~\footnote{www.cse.unsw.edu.au/$\sim$cs9323} during 2009 and 2010. There are different people (e.g. students, mentors and lecturers) involved in this course. For example in semester 2-2009 we had 66 people (60 students + 5 project mentors + 1 lecturer) involved in the course activities. During this semester, fifteen project groups (each group consists of four students) have been formed where each group has been allocated to one of the available projects. Each mentor has been allocated to supervise three projects. The development process of each project has gone through a sequence of pre-defined  phases:  brainstorming, requirements analysis, design phase, prototype implementation, testing and final product delivery.

\begin{figure}
\centering
  \includegraphics[scale=0.5]{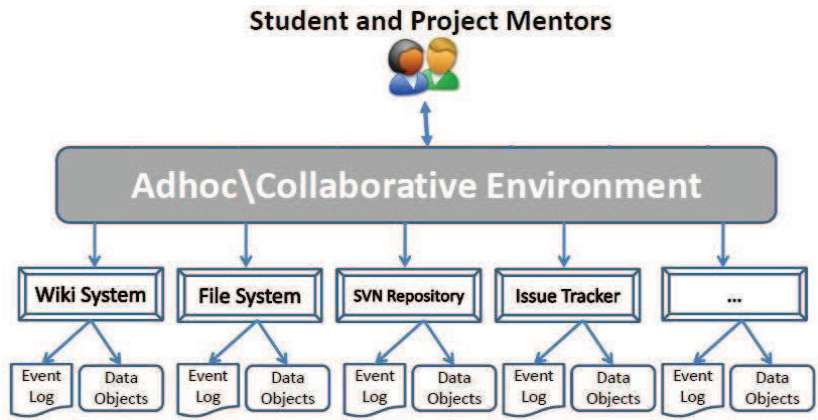}\\
  \caption{Evaluation example.}\label{MotivatingScenario}
\end{figure}

The activities of each project have been documented through a Web-based project management system which is equipped with many back-end modules such as: 1) \textsf{Message Board} to exchange message and open discussion topics between the project members. 2) \textsf{Wiki System} which is used to collaboratively edit related documents to the activities of their project. 3) \textsf{Blogging System} where each user has his own blog to edit his own posts. 4) \textsf{File Sharing System} where project members can share access to different files and documents. 5) \textsf{SVN Repository} to synchronize the editing of the projects source code. Figure~\ref{MotivatingScenario} depicts an illustration of our motivating scenario.

The graph structure in our example is made up of entities such as artifact, process, and agent. An artifact is an immutable piece of state which has a digital representation in a computer system, e.g. a file. A process is an action or series of actions performed on or caused by artifacts. An agent is an entity which is capable of acting as a catalyst of a process. These entities are connected by one or more specific types of interdependency, such as 'process used artifact', 'process \emph{wasTriggeredBy} process', 'process \emph{wasControledBy} agent', 'artifact \emph{wasGeneratedBy} process', and 'artifact \emph{wasDerivedFrom} artifact' (for more detail see~\cite{Moreau:OPM}).

\subsection{Query Execution Time} \label{QueryExecutionTime}

We evaluated the performance of the FPSPARQL query engine compared to one of the well-known graph databases, the HyperGraphDB~\cite{HyperGraphDB}. There is no query language for HyperGraphDB and querying is performed through special purpose APIs. These APIs are based on conditional expressions that a user creates, submits to the query system and receives a set of nodes as the result. We extracted and simulated over one million events (about 25 million triples) out of e-Enterprise course database (section \ref{MotivatingExample}) to generate a large RDF file (1.9 GByte). It took 22.8 minutes to load the input RDF file into FPSPARQL relational RDF store. HyperGraphDB manages storage as a set of files in a directory. To create a database and load the same file as input into HyperGraphDB, we have implemented a loader. The loader took 52.2 minutes to load the input file. In the appendix we present FPSPARQL query samples that were useful for our e-Enterprise course collaborators. For each query expressed in English, we construct a FPSPARQL query and its equivalent SQL queries, generated by FPSPARQL-to-SQL translation algorithm.

\begin{figure}
\centering
  \includegraphics[scale=0.6]{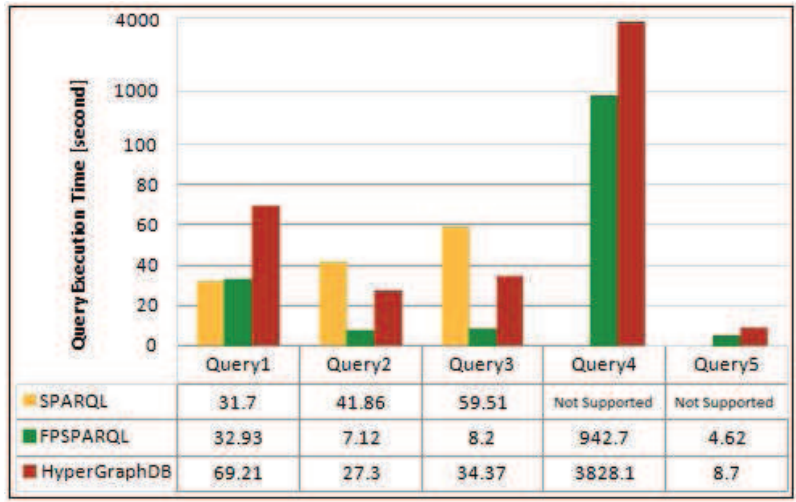}\\
  \caption{Query Execution Times.}\label{Evaluation}
\end{figure}

Figure~\ref{Evaluation} illustrates the query execution time for each FPSPARQL query, its SPARQL equivalent, and HyperGraphDB API. Query1 is a folder node construction query. Query1 runs a bit faster on SPARQL, compared to FPSPARQL. The reason could be a small overhead for storing the folder in FPSPARQL query. Both SPARQL and FPSPARQL executed faster than HyperGraphDB. Query2 is a folder node selection query. The execution time of FPSPARQL shows that applying queries on folder nodes, improves the query processing time of many complex queries. The equivalent SPARQL query should apply the condition on the whole graph which takes longer to execute. The execution time of HyperGraphDB is much better than SPARQL query, but not comparable to FPSPARQL query. Query3 is a folder node selection query. In FPSPARQL, the query applied on the composition of two constructed folder nodes. For HyperGraphDB we generate same folders as hypergraphs and applied a query on the composition of them. Figure~\ref{Evaluation} shows the better +performance of FPSPARQL compared to its equivalent SPARQL query and HyperGraphDB API.

Query4 is a path node construction query. FPSPARQL provides the ability to call external graph reachability algorithms in path node queries (see Section~\ref{ApproximationQuality}). It took 15.7 minutes, for the FPSPARQL engine, to parse the regular expressions and explore potential paths. As the result one path was discovered. HyperGraphDB has APIs providing the traversal algorithm (breadth-first or depth-first). The performance for these APIs depends on the incidence index and the efficient caching of incidence sets. We applied efficient index and caching to run query4 on HyperGraphDB. The query took 63.8 minutes to explore potential paths. As the result one path was discovered. We stored the path (manually) as a hypergraph to use in query5. Query 4 is not supported in SPARQL query language.

Query5 is a path node selection query. In FPSPARQL, the query applied on the path node constructed in query4. In HyperGraphDB, the query applied on the path generated in query4 which stored manually as a hypergraph. HyperGraphDB does not support the automatic construction and selection of paths. Also it does not provide the ability to call external traversal algorithms. Query 5 is not supported in SPARQL query language. Figure~\ref{Evaluation} illustrates the performance of these queries.

\subsection{Graph Reachability Analysis} \label{ApproximationQuality}

We developed an interface to support various graph reachability algorithms~\cite{GraphBook} such as Transitive Closure, GRIPP, Tree Cover, Chain Cover, Path-Tree Cover, and Shortest-Paths~\cite{ShortestPath}. In general, there are two types of graph reachability algorithms~\cite{GraphBook}: (1) algorithms traversing from starting vertex to ending vertex using breadth-first or depth-first search over the graph, and (2) algorithms checking whether the connection between two nodes exists in the edge transitive closure of the graph. Considering  $G=(V,E)$ as directed graph that has $n$ nodes and $m$ edges, the first approach incurs high cost as $O(n+m)$ time which requires too much time in querying. The second approach results in high storage consumption in $O(n^2)$ which requires too much space. In this experiment, we used the GRIPP~\cite{GRIPP} algorithm which has the querying time complexity of $O(m-n)$, index construction time complexity of $O(n+m)$, and index size complexity of $O(n+m)$.

\section{Related Work}
\label{RelatedWork}

A recent book~\cite{GraphBook} and survey \cite{SurveyGraph} discuss a number of data models and query languages for graph data. Some of existing approach for querying and modeling graphs~\cite{QueryLangAnalyzNet,InformationFragments} focused on defining constraints on nodes and edges simultaneously on the entire object of interest, not in an iterative one-node-at-a-time manner. Therefore, they do not support querying nodes at highler levels of abstraction. Authors of~\cite{InformationFragments} propose an \emph{Information Fragment} as an abstraction for representing a subgraph. They do not support querying information fragments.

BiQL~\cite{QueryLangAnalyzNet} is an SQL-based query language focused on the uniform treatment of nodes and edges and supports queries that return subgraphs. BiQL supports a closure property on the result of its queries meaning that the output of every query can be used for further querying. Compared to BiQL, in our work folders and paths are first class abstractions (graph nodes) and can be defined in a hierarchical manner, over which queries are supported. 

HyperGraphDB \cite{HyperGraphDB} is a graph database based on hypergraphs (a hypergraph node is connected through an edge to all vertices that are contained in it). There is no query language for HyperGraphDB and querying is performed through special purpose APIs. HyperGraphDB builds on two prior approaches of Hypernode~\cite{hypernodeModel} and GROOVY~\cite{GROOVY} graph representation models which focus on representing objects and object schemas. 
GROOVY \cite{GROOVY} and Hypernode only support typed objects, and have no support for hypernode specific operations.

SPARQL \cite{SPARQL} is a declarative query language, an W3C standard, for querying and extracting information from directed labeled RDF graphs. SPARQL supports queries consisting of triple patterns, conjunctions, disjunctions, and other optional patterns. However, there is no support for querying grouped entities. Paths are not first class objects in SPARQL~\cite{SPARQL,ProvQuery}. PSPARQL \cite{PSPARQL} extends SPARQL with regular expressions patterns allowing path queries. SPARQLeR \cite{SPARQLeR} is an extension of SPARQL designed for finding semantic associations (and path patterns) in RDF bases. 
In FPSPARQL, we support folder and path nodes as first class entities that can be defined at several levels of abstractions and queried. In addition, we provide an efficient implementation of a query engine that support their querying.

\section{Conclusion} \label{Conclusion}

In this paper, we presented a data model and query language for querying and analyzing large graphs specifically for analyzing groups of related entities. The data model supports structured and unstructured entities, and introduces folder and path nodes as first class abstractions. The query language, i.e. FPSPARQL, defined as an extension of SPARQL to manipulate and query entities, and folder and path nodes. We have developed an efficient and scalable implementation of FPSPARQL within the high-performance relational RDF storage and retrieval system. To evaluate the viability and efficiency of FPSPARQL, we have conducted experiments over large graph datasets. We compared the quality and speed of FPSPARQL with HyperGraphDB~\cite{HyperGraphDB}.

As future work, we plan to design a visual query interface to support users in expressing their queries over the conceptual representation of the graph in an easy way. Moreover, we plan to make use of interactive graph exploration and visualization techniques (e.g. storytelling systems~\cite{Tolkien}) which can help users to quickly identify the interesting parts of a graph. We are also interested in the temporal aspects of graph analysis, as in some cases (e.g. provenance) the structure of the graph may change rapidly over time.

\bibliographystyle{plain}
\bibliography{mybib}

\begin{thebibliography}{10}

\bibitem{GraphBook}
Charu~C. Aggarwal and Haixun Wang.
\newblock {\em Managing and Mining Graph Data}.
\newblock Springer Publishing Company, Incorporated, 2010.

\bibitem{RE_path}
Faisal Alkhateeb, Jean-Fran\c{c}ois Baget, and J{\'e}r{\^o}me Euzenat.
\newblock Extending sparql with regular expression patterns (for querying rdf).
\newblock {\em J. Web Sem.}, 7(2):57--73, 2009.

\bibitem{SurveyGraph}
Renzo Angles and Claudio Gutierrez.
\newblock Survey of graph database models.
\newblock {\em ACM Comput. Surv.}, 40:1:1--1:39, February 2008.

\bibitem{PSPARQL}
Kemafor Anyanwu, Angela Maduko, and Amit Sheth.
\newblock Sparq2l: towards support for subgraph extraction queries in rdf
  databases.
\newblock In {\em WWW '07: Proceedings of the 16th international conference on
  World Wide Web}, pages 797--806, New York, NY, USA, 2007. ACM.

\bibitem{CodebookPath}
Andrew Begel, Yit~Phang Khoo, and Thomas Zimmermann.
\newblock Codebook: discovering and exploiting relationships in software
  repositories.
\newblock In Jeff Kramer, Judith Bishop, Premkumar~T. Devanbu, and
  Sebasti{\'a}n Uchitel, editors, {\em ICSE (1)}, pages 125--134. ACM, 2010.

\bibitem{RDFProv}
Artem Chebotko, Shiyong Lu, Xubo Fei, and Farshad Fotouhi.
\newblock Rdfprov: A relational rdf store for querying and managing scientific
  workflow provenance.
\newblock {\em Data Knowl. Eng.}, 69(8):836--865, 2010.

\bibitem{SPARQLSemantics}
Artem Chebotko, Shiyong Lu, and Farshad Fotouhi.
\newblock Semantics preserving sparql-to-sql translation.
\newblock {\em Data Knowl. Eng.}, 68(10):973--1000, 2009.

\bibitem{SPARQLAlgebra}
Richard Cyganiak.
\newblock A relational algebra for {SPARQL}.
\newblock 2005.

\bibitem{QueryLangAnalyzNet}
Anton Dries, Siegfried Nijssen, and Luc De~Raedt.
\newblock A query language for analyzing networks.
\newblock In {\em CIKM '09: Proceeding of the 18th ACM conference on
  Information and knowledge management}, pages 485--494, New York, NY, USA,
  2009. ACM.

\bibitem{InformationFragments}
Thomas Fritz and Gail~C. Murphy.
\newblock Using information fragments to answer the questions developers ask.
\newblock In {\em ICSE '10: Proceedings of the 32nd ACM/IEEE International
  Conference on Software Engineering}, pages 175--184, New York, NY, USA, 2010.
  ACM.

\bibitem{ShortestPath}
Andrey Gubichev, Srikanta~J. Bedathur, Stephan Seufert, and Gerhard Weikum.
\newblock Fast and accurate estimation of shortest paths in large graphs.
\newblock In {\em CIKM}, pages 499--508, 2010.

\bibitem{ProvQuery}
David~A. Holl, Uri Braun, Diana Maclean, Kiran kumar Muniswamy-reddy, and
  Margo~I. Seltzer.
\newblock Choosing a data model and query language for provenance, 2008.

\bibitem{HyperGraphDB}
Borislav Iordanov.
\newblock Hypergraphdb: A generalized graph database.
\newblock In {\em WAIM Workshops}, pages 25--36, 2010.

\bibitem{SPARQLeR}
Krys~J. Kochut and Maciej Janik.
\newblock Sparqler: Extended sparql for semantic association discovery.
\newblock In {\em Proceedings of the 4th European conference on The Semantic
  Web: Research and Applications}, ESWC '07, pages 145--159, Berlin,
  Heidelberg, 2007. Springer-Verlag.

\bibitem{GROOVY}
M.~Levene and A.~Poulovanssilis.
\newblock An object-oriented data model formalised through hypergraphs.
\newblock {\em Data Knowl. Eng.}, 6:205--224, May 1991.

\bibitem{hypernodeModel}
M.~Levene and A.~Poulovassilis.
\newblock The hypernode model and its associated query language.
\newblock In {\em Proceedings of the fifth Jerusalem conference on Information
  technology}, JCIT, pages 520--530, Los Alamitos, CA, USA, 1990. IEEE Computer
  Society Press.

\bibitem{Moreau:OPM}
Luc Moreau, Juliana Freire, Joe Futrelle, Robert~E. Mcgrath, Jim Myers, and
  Patrick Paulson.
\newblock The open provenance model: An overview.
\newblock pages 323--326, 2008.

\bibitem{OneTableStoresAll}
Beng~Chin Ooi, Bei Yu, and Guoliang Li.
\newblock One table stores all: Enabling painless free-and-easy data publishing
  and sharing.
\newblock In {\em CIDR}, pages 142--153, 2007.

\bibitem{SPARQL}
Eric Prud'hommeaux and Andy Seaborne.
\newblock Sparql query language for rdf (working draft).
\newblock Technical report, W3C, March 2007.

\bibitem{SherifSurvey1}
Sherif Sakr and Ghazi Al-Naymat.
\newblock Relational processing of rdf queries: a survey.
\newblock {\em SIGMOD Rec.}, 38(4):23--28, 2009.

\bibitem{Tolkien}
Arjun Satish, Ramesh Jain, and Amarnath Gupta.
\newblock Tolkien: an event based storytelling system.
\newblock {\em Proc. VLDB Endow.}, 2:1630--1633, August 2009.

\bibitem{GRIPP}
Silke TriBl and Ulf Leser.
\newblock Fast and practical indexing and querying of very large graphs.
\newblock In {\em Proceedings of the 2007 ACM SIGMOD international conference
  on Management of data}, SIGMOD '07, pages 845--856, New York, NY, USA, 2007.
  ACM.

\end{thebibliography}

\section*{Appendix: FPSPARQL Queries}
\label{Appendix}

In this section we present FPSPARQL query samples that were useful for our e-Enterprise course collaborators. For each query expressed
in English, we construct a FPSPARQL query and its equivalent SQL query, generated by FPSPARQL-to-SQL translation algorithm.\\\\

\noindent\textbf{Query 1.} [Folder Node Construction] Group all the events happened in the context of \emph{brainstorming} phase during "semester 2, 2009", in a folder named "brainstorming09s2". Brainstorming phase \emph{start time} is '19 July 2009' and \emph{end time} is '8 August 2009'.\\

\noindent FPSPARQL:
\begin{verbatim}
  fconstruct brainstorming09s2 as ?fn
  select ?e
  where{
  ?fn @description 'related events...'.
  ?e @type Event.
  ?e @timestamp ?date.
  FILTER (?date > "2009-07-19" ^^xsd:date &&
  ?date > "2009-08-08" ^^xsd:date).
  }
\end{verbatim}

\noindent \\SQL:
\begin{verbatim}
  @folderID <- generate a unique folderID
  ...

  insert into NULLID.EntityStore
  (subject , predicate , object)
  values
  (@folderID , '@Name' , 'brainstorming09s2');
  insert into NULLID.FOLDERSTORE
  (folderid , subject , predicate , object)
  select @folderID as folderid , subject , predicate , object
  from NULLID.GraphStore
  where subject in
  (SELECT r.e AS e
  FROM (
  Select es1.subject AS e, es2.object AS date
  From NULLID.EntityStore es1, NULLID.EntityStore es2
  Where es1.predicate = '@type' AND es1.object = 'Event' AND
  es2.predicate = '@timestamp' AND es1.subject = es2.subject  AND
  (DATE(SUBSTRING(es2.object,1,10,CODEUNITS32)) > DATE(2009-07-19)
  AND
  DATE(SUBSTRING(es2.object,1,10,CODEUNITS32)) > DATE(2009-08-08))
  ) AS r );
\end{verbatim}

\begin{center}
\line(1,0){250}
\end{center}

\noindent\textbf{Query 2.} [Folder Node Selection] Return the list of artifacts that have been part of update events which are triggered by a comment event in the context of \emph{brainstorming} phase during "semester 2, 2009", i.e. the folder we created in Query1.\\

\noindent FPSPARQL:
\begin{verbatim}
  (brainstorming09S2) apply (
  select ?a
  where {
  ?e @type 'Event'.
  ?e @activityType 'update'.
  ?e @ArtifactName ?a.
  ?e wasTriggeredBy ?x.
  ?x @type 'Event'.
  ?x @activityType 'comment'.
  })
\end{verbatim}

\noindent \\SQL:
\begin{verbatim}
  SELECT r.a AS a FROM (
  Select es1.subject AS e, es3.object AS a, fs1.object AS x
  From NULLID.EntityStore es1, NULLID.EntityStore es2,
  NULLID.EntityStore es3, NULLID.FOLDERSTORE fs1,
  NULLID.EntityStore es4, NULLID.EntityStore es5
  Where es1.predicate = '@type' AND es1.object = 'Event'
  AND es4.object = 'Event' AND es2.predicate = '@activityType'
  AND es2.object = 'update' AND es3.predicate = '@ArtifactName'
  AND fs1.predicate in (
  select subject
  from NULLID.EntityStore
  where predicate = '@Label' AND object = 'wasTriggeredBy' )
  AND fs1.FolderID in (
  Select subject
  from NULLID.EntityStore
  where predicate = '@Name' AND object = 'brainstorming09S2')
  AND es5.object = 'comment' AND es1.subject = es2.subject
  AND es1.subject = es3.subject AND es1.subject = fs1.subject
  AND es1.object = es4.object AND es4.subject = es5.subject
  AND fs1.object = es4.subject AND es1.predicate=es4.predicate
  AND es2.predicate = es5.predicate ) AS r
\end{verbatim}

\begin{center}
\line(1,0){250}
\end{center}

\noindent\textbf{Query 3.} [Folder Node Selection] Return the list of users who were involved in updating an artifact during \emph{brainstorming} and \emph{design} phase of semester 1, 2010. We construct two folders of all events that happened in the context of brainstorming phase (brainstorming10s1), and design phase (design10s1) during "semester 1 2010".\\

\noindent FPSPARQL:
\begin{verbatim}
  (brainstorming10s1 union design10s1) apply (
  select ?u
  where {
  ?e @type 'Event'.
  ?e @activityType 'update'.
  ?e @UseName ?u.
  })
\end{verbatim}

\noindent \\SQL:
\begin{verbatim}
  SELECT r.u AS u FROM (
  Select es1.subject AS e, es3.object AS u
  From NULLID.EntityStore es1, NULLID.EntityStore es2,
  NULLID.EntityStore es3
  Where es1.predicate = '@type' AND es1.object = 'Event' AND
  es2.predicate = '@activityType' AND es2.object = 'update' AND
  es3.predicate = '@UseName' AND es1.subject = es2.subject AND
  es1.subject = es3.subject AND es1.subject in (
  select subject
  from NULLID.FOLDERSTORE
  where folderid in (
  select subject from NULLID.EntityStore
  where predicate = '@name' and subject = 'brainstorming10s1')
  union
  select subject
  from NULLID.FOLDERSTORE
  where folderId in (
  select subject
  from NULLID.EntityStore
  where predicate = '@name' and subject = 'design10s1') ) ) AS r
\end{verbatim}

\begin{center}
\line(1,0){250}
\end{center}


\noindent\textbf{Query 4.} [Path Node Construction] Construct a path between the event that generates brainstorming document (brainDoc.doc), and the event that generates design document (designDoc.doc) which were rendered by project4 members during semester 2, 2009. This path should contain the pattern of an event responding to a bug report in the Wiki.\\

\noindent FPSPARQL:
\begin{verbatim}
  pconstruct myPathNode
  (?startNode,?endNode,(?e ?n)* ?e ?node ?e (?n ?e)* )
  where {
  ?startNode @type 'Event'.
  ?startNode @activityType 'generate'.
  ?startNode @artifactName 'brainDoc.doc'.
  ?startNode @UserGroup 'project4'.
  ?startNode @timestamp ?date.
  ?endNode @type 'Event'.
  ?endNode @activityType 'generate'.
  ?endNode @artifactName 'designDoc.doc'.
  ?endNode @UserGroup 'project4'.
  ?endNode @timestamp ?date.
  ?n @isA 'entityNode'.
  ?n @type 'Event'.
  ?n @timestamp ?date.
  ?e @isA 'edge'.
  ?node @type 'Event'.
  ?node @activityType 'response'.
  ?node @layer 'Wiki'.
  ?node @layerPart 'bug'.
  ?node @timestamp ?date.
  FILTER (?date > "2009-07-19" ^^xsd:date &&
  ?date > "2009-11-04" ^^xsd:date). }
\end{verbatim}

\noindent \\SQL:
\begin{verbatim}
  A graph reachability algorithm used (see section 5).
\end{verbatim}

\begin{center}
\line(1,0){250}
\end{center}

\noindent\textbf{Query 5.} [Path Node Selection] Return the list of artifacts that generated between the path constructed in Query4.\\

\noindent FPSPARQL:
\begin{verbatim}
  (myPathNode) apply (
  select ?a
  where {
  ?e @type 'Event'.
  ?e @activityType 'generate'.
  ?e @ArtifactName ?a.
  })
\end{verbatim}

\noindent \\SQL:
\begin{verbatim}
  SELECT r.a AS a FROM (
  Select es1.subject AS e, es3.object AS a
  From NULLID.EntityStore es1, NULLID.EntityStore es2,
  NULLID.EntityStore es3
  Where es1.predicate = '@type' AND es1.object = 'Event' AND
  es2.predicate = '@activityType' AND es2.object = 'generate'
  AND es3.predicate='@ArtifactName' AND es1.subject=es2.subject
  AND es1.subject = es3.subject AND es1.subject in (
  SELECT subject
  from NULLID.PathStore
  Where PathId in (
  SELECT subject
  from NULLID.EntityStore
  Where predicate = '@name' AND object='myPathNode'))) AS r
\end{verbatim}

\begin{center}
\line(1,0){250}
\end{center}

\end{document}